\documentclass[a4,center,fleqn,twocolumn]{article}
\usepackage[utf8]{inputenc}
\usepackage{color}
\usepackage{graphicx}
\usepackage{subfigure}
\usepackage{amsmath}
\usepackage{url}
\usepackage{cite}

\begin{document}

\title{Probe region expression estimation for RNA-seq data for
  improved microarray comparability}
 
\author{%
  Karolis Uziela\,$^{1,2}$
  and
  Antti Honkela\,$^{1}$\\[1em]
  \small $^1$ Helsinki Institute for Information Technology HIIT,
  Department of Computer Science, \\
  \small University of Helsinki, Helsinki, Finland \\
  \small $^2$ Science for Life Laboratory and Department of Biochemistry and
  Biophysics, \\
  \small Stockholm University, Solna, Sweden
}%

\date{}

\maketitle

\begin{abstract}
  Rapidly growing public gene expression databases contain a wealth of
  data for building an unprecedentedly detailed picture of human
  biology and disease.  This data comes from many diverse measurement
  platforms that make integrating it all difficult.
  Although RNA-sequencing (RNA-seq) is attracting the most attention,
  at present the rate of new microarray studies submitted to public
  databases far exceeds the rate of new RNA-seq studies.
  There is clearly a need for methods that make it easier to combine
  data from different technologies.
  In this paper, we
  propose a new method for processing RNA-seq data that yields
  gene expression estimates that are much more similar to
  corresponding estimates from microarray data, hence greatly
  improving cross-platform comparability.  The method we call PREBS is
  based on estimating the expression from RNA-seq reads
  overlapping the microarray probe
  regions, and processing these estimates with microarray
  summarisation algorithm RMA.  
  Using paired microarray and RNA-seq samples from TCGA LAML data set we
  show that PREBS expression estimates derived from RNA-seq are
  more similar to microarray-based expression estimates than
  those from other RNA-seq processing methods.
  In an experiment to retrieve paired microarray samples from a
  database using an RNA-seq query sample,
  gene signatures defined based on PREBS expression estimates
  were found to be much more accurate than those from other methods.
  PREBS also allows new ways of using
  RNA-seq data, such as expression estimation for microarray
  probe sets.
  An implementation of
  the proposed method is available in the Bioconductor package `prebs'.
\end{abstract}


\section{Introduction}

Public gene expression databases such as
ArrayExpress~\cite{Brazma2003} and Gene Expression
Omnibus~\cite{Edgar2002} host public data from more than half a
million gene expression experiments.  While the field is moving toward
sequencing-based methods for expression analysis, an overwhelming
majority of the existing and even newly uploaded data in these
databases are still from
microarray platforms as demonstrated in Table~\ref{table:platform_stat}.
The existing microarray-based data represent a
huge investment and being able to utilise it efficiently as background
information in new sequencing-based studies is of great interest.

\begin{table*}[t]
\begin{tabular}{|l|l|l|l|l|l|l|} 
\hline
& \textbf{2010} & \textbf{2011} & \textbf{2012} & \textbf{2013} & \textbf{2014 (extrapolated)} & \textbf{All time} \\ 
\hline 
\textbf{RNA-seq} & 243 & 449 & 779 & 1339 & 1881 & 4063 \\ 
\hline
\textbf{Microarray} & 6034 & 5605 & 6052 & 6536 & 6139 & 38029 \\ 
\hline
\end{tabular}
\caption{Number of RNA-seq and microarray experiments in ArrayExpress database as of August 1, 2014}
\label{table:platform_stat}
\end{table*}

Recently there has been significant interest in utilising the large
public databases to holistically characterise phenotypes based on
expression in new samples~\cite{Schmid2012}.  Most work utilising
these large databases is based on differential
expression~\cite{Caldas2009,Huang2010,Caldas2012a}, but Schmid
\emph{et al.}~\cite{Schmid2012} argue that absolute expression can
yield a more
comprehensive picture.  All of these methods are
currently restricted to microarray data, which severely limits their
utility in new studies.

RNA-seq and microarrays are based on very different principles and
ultimately measure different things~\cite{Malone2011}.  Numerous
experimental comparisons have demonstrated RNA-seq and microarrays to
yield broadly comparable
results~\cite{marioni2008,Fu2009,Bradford2010,Su2011,Bottomly2011,beane2011,Nookaew2012,Arino2013}.
These results demonstrate that the platforms typically agree on
differentially expressed genes between sufficiently different samples,
although RNA-seq tends to be more sensitive.  For measures of absolute
expression, there is typically a clear correlation, the level of which
ranges from moderate to very high depending on the example.

In this paper we present a method for processing RNA-seq data in a way
to make the resulting expression measures significantly more
comparable with measures derived from microarray data by estimating
the expression level at the microarray probe regions using a method
we call PREBS (Probe Region Expression estimation Based on Sequencing).  The
improvement is especially significant for measures of absolute
expression.  This improved comparability comes at the expense of
ignoring some information in the RNA-seq data by focusing the analysis
to regions covered by the microarray probes.  Because of this loss of
information, PREBS should not be viewed as a replacement of standard
RNA-seq analysis tools.  Neither is it a replacement for actually
performing the corresponding microarray experiment if the sample
material and sufficient resources are available, but rather a cheap
computational alternative for the very common case when either samples
or resources are not available.

\section{MATERIALS AND METHODS}

\subsection{Basic description of the method}

One of the fundamental differences between microarray and RNA-seq technologies
is that microarrays, especially now ubiquitous oligonucleotide arrays,
measure gene expression based on the parts of the
gene where probe sequences are located~\cite{Lockhart2000}
while RNA-seq measures expression over the whole gene
sequence~\cite{mortazavi2008}. The idea of our method is to
eliminate this difference by calculating RNA-seq gene expression measures only
based on the parts of the gene where microarray probe sequences are
located.

Traditionally gene expression is estimated from RNA-seq data by
counting the number of reads that overlap with exons of the gene
(count methods)~\cite{mortazavi2008,wang2009}.  The analysis in higher
eukaryotes can be complicated by alternative splicing.  To account for
this, several methods have been proposed that are based on deconvolution of
transcript isoform expression using probabilistic
models~\cite{Trapnell2010,Li2010e,turro2011,Glaus2012}, but these
methods still estimate the expression level across the whole gene.

In PREBS method we estimate probe region expression by
counting the number of reads that overlap with probe regions and using
a statistical model to infer the expression level from the read
counts. We treat the inferred probe region expression levels
in a similar way as they are treated in computational
microarray processing pipelines. In particular, we
apply the popular microarray data summarisation
algorithm RMA~\cite{irizarry2003} commonly used for Affymetrix data
analysis. The details of the RMA algorithm
application and the statistical model used to infer probe regions will
also be described in later sections.

Using the described method we aim to computationally process RNA-seq
data in a way that is similar to microarray computational processing
pipelines. In the Results section we show that gene expression
measures that we get from RNA-seq data this way are more similar to
microarray measures than the measurements that we get using
conventional RNA-seq data processing methods. We call our RNA-seq
data processing method PREBS (Probe Region Expression estimation Based on
Sequencing).

\subsection{Read counting}

For counting read overlaps PREBS uses count\_overlaps() function from
GenomicRanges package in R/Bioconductor. Just like implemented in
count\_overlaps() function, PREBS counts the read for all overlapping
probe regions, even if one read overlaps with several of them. There
is no need to discard reads that overlap several probe regions,
because it would cause biased under-expression of densely probe-packed
genome areas. Moreover, PREBS has inherited a feature from
count\_overlaps() function that allows to select whether the strand
from which the read originates should be ignored when counting the
overlaps. Since most of the RNA-seq protocols that are used nowadays
are not strand-specific, the default behavior of PREBS is to ignore
the strand. Finally, PREBS supports a possibility to process both
single-ended and paired-ended reads. If paired-ended mode is selected,
the two mates are treated as a single unit, not as independent reads
during read-counting process.

\subsection{Probe region expression estimation from RNA-seq}

Read sampling in sequencing is inherently a stochastic process.  To
account for the uncertainty this induces, we use statistical methods
to infer the probe region expression level from read data.

We assume that the number of reads from a region with a given
expression level follows the Poisson distribution.
Placing a conjugate gamma prior on the expression level, we obtain an
estimate of the expression level as the mean of the posterior
distribution. The hyperparameters of the prior are determined using an
empirical Bayesian approach by maximising the marginal likelihood of
the full data.

\subsection{Expression summarisation by PREBS-RMA}

Affymetrix microarray probes are grouped into probe sets containing
15-20 perfect match / mismatch probe pairs. Perfect match probes are
completely complementary to gene portion they are interrogating while
mismatch probes have their middle nucleotide changed. Some algorithms
like MAS5~\cite{mas5} use expression values from mismatch probes to
account for non-specific binding while RMA completely ignores mismatch
probe values and uses only the perfect match probes.

We used a modified version of the rma() function in the
R/Bioconductor Affy package~\cite{Gautier2004} to apply the RMA
algorithm to probe region expression estimates. 
The noise characteristics of microarrays and RNA-seq are different,
especially at the low end of the expression level spectrum, where
microarrays have a significant background that is removed by the
background correction step in the RMA algorithm.  This step is not
useful for RNA-seq data, so we omitted it.
Since
RMA algorithm normalises all samples at the same
time~\cite{irizarry2003}, custom microarray probe expression values
have to be supplied to the function from all of the samples
simultaneously. The two other major steps of RMA algorithm,
normalisation and summarisation, were left unchanged and performed as
they are implemented in Affy package.

When processing microarray data using RMA algorithm from the Affy
package, the user has two options: process the data based on original
microarray probe set definitions or based on alternative probe
set definitions using so called Custom CDF
files~\cite{dai2005}. By default, the resulting expression
values are calculated for the original microarray probe sets. On the
other hand, when the data are processed using Custom CDF files, the
expression measures can be directly calculated for other units such as
Ensembl genes. The latter option greatly simplifies the comparison
between microarray and RNA-seq data, since microarray gene expression
values calculated for Ensembl genes can be directly compared with
the gene expression values calulated using various RNA-seq data
processing tools.

PREBS shares the feature of being able to run in the same two modes.
On the one hand, the values that we get using Custom CDF files
for Ensembl genes can be easily compared with RNA-seq
gene expression values and therefore, most of the results in this
paper are based on this mode. On the other hand, being able to
get expression values for the original
probe sets is a unique feature of PREBS that no other RNA-seq data
processing method possesses. This feature is certainly very useful for
people who prefer to work on expression summaries for microarray probe
sets but still want to compare these to RNA-seq expression estimates.

\subsection{Tools used for implementation}

In order to evaluate the effectiveness of our method (PREBS) we
compared it to representatives of two RNA-seq analysis
methods: count-based~\cite{mortazavi2008} (``Read counting'')
and isoform deconvolution (``MMSEQ'').
We processed
sequencing data using each of the methods and evaluated their
agreement with microarray data by calculating correlations of gene
expression.

For the PREBS method, reads were mapped by TopHat software version
1.4.1~\cite{trapnell2009} to Human genome version GRCh37.65.
We considered only unique genomic alignments to annotated
transcripts. When running PREBS with Ensembl gene summaries, the locations for probe regions were retrieved from Custom CDF file
annotations (version 15.0.0 ENSG)~\cite{dai2005}. For probe set
summaries, we mapped the probe sequences to the latest human genome
build (hg19) using Bowtie (version 0.12.7). The read overlaps with probe regions
were calculated using GenomicRanges package from
R/Bioconductor~\cite{gentleman2004}. Probe region expression estimates were
calculated as described above and fed to a modified version of the
rma() function from R/Bioconductor Affy package.

Read counting RPKM values were calculated using the same tools as in PREBS method,
but read overlap counts were calculated for Ensembl genomic
annotations that were downloaded using GenomicFeatures package. RPKM
values were calculated using these counts and $\log_2$ values
were taken.

For isoform deconvolution we used MMSEQ~\cite{turro2011}
(software version 0.9.18).
Bowtie software (version 0.12.7)~\cite{langmead2009} was
used to map the reads to the transcriptome, as recommended by MMSEQ
manual. MMSEQ options were set to default and Bowtie options were set
as recommended by MMSEQ (-a --best --strata -S -m 100 -X 400).
Human transcriptome version GRCh37.65 from Ensembl database was
used. MMSEQ output values were converted from natural logarithm scale to $\log_2$ scale.

Microarray expression values were summarised using
rma() function from Affy package. In case of multiple replicates, the mean value was taken as an absolute expression estimate for each state. RMA-summarised values were in $\log_2$ scale, so no logarithm base conversion was needed.

Significance tests between the observed correlation differences were
performed using r.test() function from psych package in R.

\section{RESULTS}

\subsection{Data sets}

\begin{figure*}[ht]
  \centering
  \includegraphics[width=\textwidth]{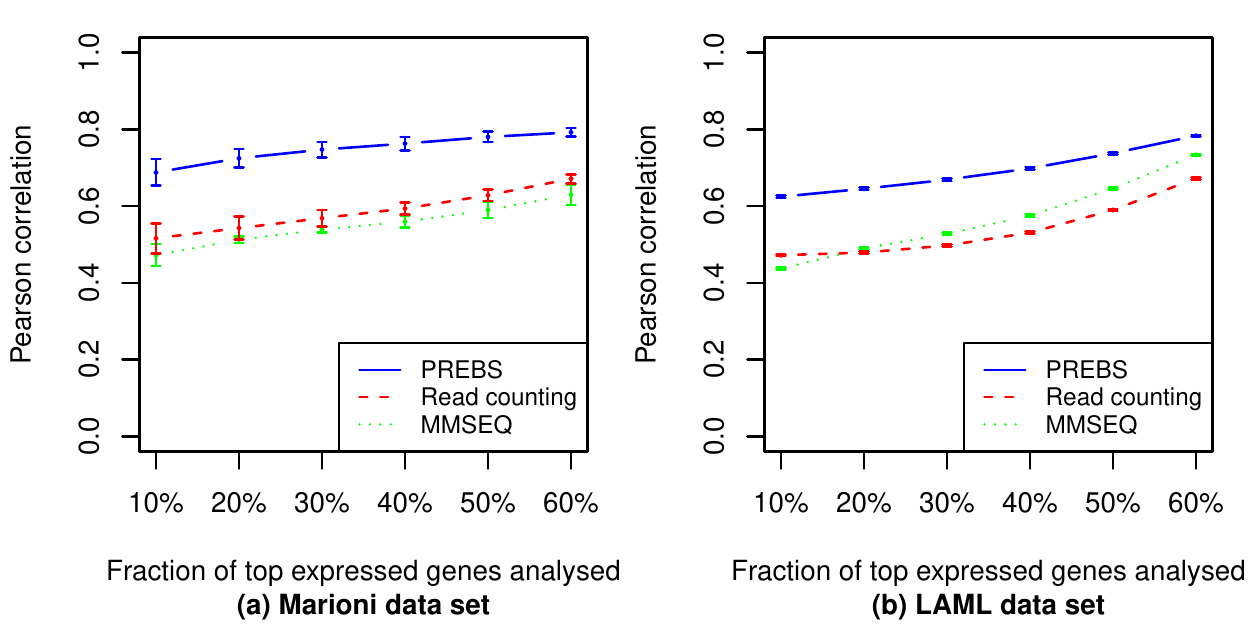}
  \caption{{\bf Averaged absolute gene expression correlations.} The plots show average absolute gene expression correlations between different RNA-seq
    data processing methods and the microarray. Different points correspond to different numbers of top expressed genes. The correlations are averaged over all samples
    in the corresponding data sets: (a) the Marioni \emph{et al.} data set,
    (b) the LAML data set. The error bars correspond to standard
    errors of the mean. For LAML data set the standard errors are so
    small that the top and bottom error bars are merged in the plot. }
  \label{fig:corplots1}
\end{figure*}

\begin{figure*}[ht] 
 \centering 
 \includegraphics[width=\textwidth]{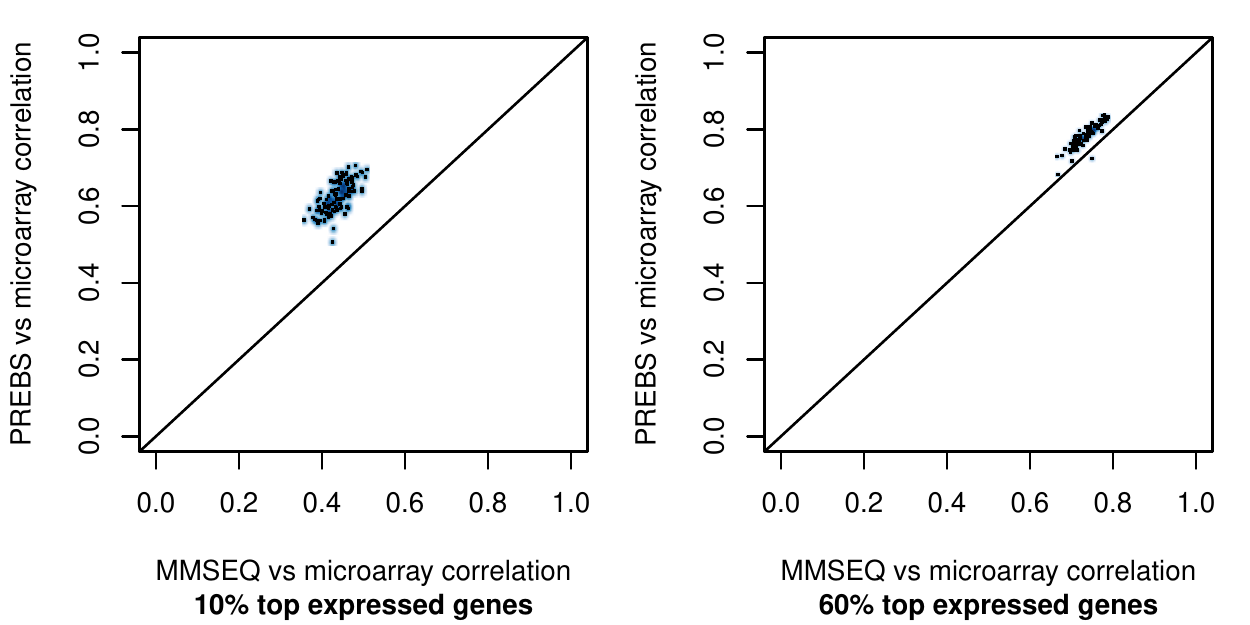}
  
  \caption{{\bf Absolute gene expression correlation scatter
      plots.} The plots show the comparison of correlations of PREBS
    vs microarray and MMSEQ vs microarray for all of the samples in
    the LAML data set. Each point represents one sample. Two different
    percentages of top expressed genes are taken: (a) 10\%, (b) 60\%.}
  \label{fig:corscatter}
\end{figure*}

\begin{figure*}[ht] 
 \centering 
 \includegraphics[width=\textwidth]{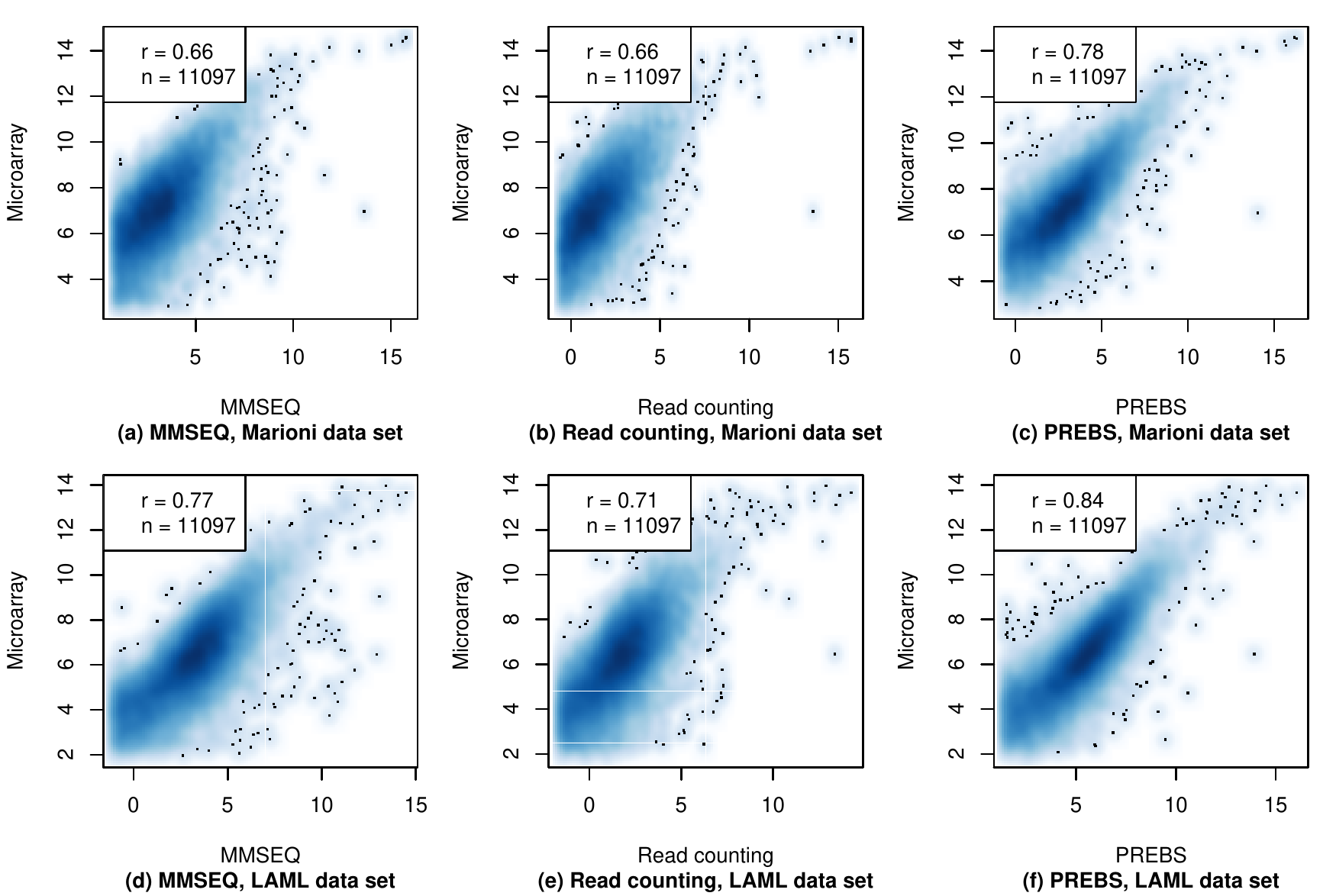}
  \caption{{\bf Absolute gene expression scatter plots.} The gene expression values from three different RNA-seq data processing methods (MMSEQ, Read counting and PREBS) are plotted against gene expression values from microarray. Only plots for a single sample in each data set are shown. The top row shows results for the
    kidney sample from the Marioni \textit{et al.}\ data set and the
    bottom row for the 2803 sample from the LAML data set. The figures show 60\%
    of most highly expressed genes. The legend
    contains Pearson correlation (\textit{r}) and the number of genes (\textit{n}).}
  \label{fig:absolute_60}
\end{figure*}

We evaluated the performance of our method on two data sets:
Marioni~\cite{marioni2008} and Acute myeloid leukaemia (LAML) from The Cancer
Genome Atlas (TCGA) database~\cite{TCGALAML2013}. These will be referred to as
Marioni and LAML data sets, respectively.
Both of these data sets have paired
RNA-seq and microarray data. The Marioni data set has two samples, human
kidney and liver, both of which were used for testing. The LAML data
set has 200 samples in all, 163 of which have both microarray and
RNA-seq data available. For 16 of those read mapping using TopHat
failed to complete (sample numbers: 2808, 2813, 2823, 2824, 2844,
2853,
2865, 2868, 2888, 2892, 2912, 2917, 2959, 2973, 2980, 2982) so we skipped
those samples and used the remaining 147 samples. In both of the data sets
RNA-seq
platform is Illumina Genome Analyzer II and microarray platform is
Affymetrix U133 Plus 2.

The main criterion for selecting the data sets was availability of
both RNA-seq and microarray data for exactly the same samples that
would be prepared in the same way. There are very few data sets
meeting this criterion. In the other data sets that had paired
RNA-seq--microarray data either the samples were different or not
prepared in the same way, or they had some other technical problems
(raw data not available, the pairing of the samples is not clear
etc.).

We were interested in checking how much information is lost in each
RNA-seq data set by only focusing on microarray probe locations in the
PREBS method. For that we computed ratios of how many of the total
reads were mapped to gene regions and out of those how many were
mapped to the microarray probe locations.
In the Marioni data set on average 79.4\% of the reads where mapped to
gene regions. Out of these, 21.1\% where mapped to microarray probe
locations inside the gene regions. In the LAML data set on average 59.1\%
reads were mapped to gene regions and 25.2\% of these were mapped to
probe locations.

\subsection{Absolute expression comparison}

First, we will present results for expression summaries for Ensembl
genes both from microarray data and PREBS.
This ensures a fair comparison against the other
RNA-seq data processing methods, as the methods we tested are able to
calculate expression values for Ensembl genes, too. The other two
RNA-seq data processing methods that PREBS was compared to were
count-based~\cite{mortazavi2008} (``Read counting'') and isoform
deconvolution (``MMSEQ'')~\cite{turro2011}.

In most gene expression studies, low expressed genes are filtered out,
because their measurements are noisy and unreliable. Common filtering
thresholds for RNA-seq data vary around 0.3 RPKM~\cite{Ramskoeld2009}.
This fraction accounts for 70\% of top expressed genes in the Marioni data set and
60.9\% of top expressed genes in the LAML data set. To make the filtering
uniform among all of the data sets and methods, we have decided to use at most
60\% of top expressed genes.

In order to evaluate the agreement of each RNA-seq data processing method
(PREBS, MMSEQ and Read counting) with microarrays, we have calculated
the Pearson correlation of sequencing-based expression values with
microarray expression values for each sample in Marioni and LAML data
sets. The correlations were calculated for different fractions of most
highly expressed genes in a sample: 10-60\%. To evaluate the methods
performance for whole data sets, we took an average correlation over
all samples in each data set (2 samples in the Marioni data set and 147
samples in the LAML data set). We provide the resulting graph that shows
the average Pearson correlations plotted as a function of the fraction
of most highly expressed genes (Figure~\ref{fig:corplots1}).

From Figure~\ref{fig:corplots1} we can clearly see that PREBS has the
best agreement with microarrays for any number of top expressed genes
taken in both data sets. The error bars that represent standard error of the mean are rather small (especially for LAML data set) which suggests that observed differences in correlation are significant. Moreover, we observe that the difference is larger for
smaller fractions of top expressed genes taken. This suggests that
PREBS is especially useful when focusing on highly expressed genes.

In order to show that the difference in correlations is robust among
different samples, we provide correlation scatter plots
(Figure~\ref{fig:corscatter}). Each point in the plot represents the
comparison of correlation between PREBS vs microarray and MMSEQ vs
microarray for a single sample in the LAML data set (so there are 147
points in each of the plots). PREBS correlation with microarray is
better than MMSEQ correlation with microarray for all of the points
that are above the diagonal. From these plots we can see that PREBS
agreement with microarray is consistently better than MMSEQ among different samples in the
LAML data set. Moreover, we can see again that the difference in
performance is larger when we take only 10\% of top expressed genes.

To give an example of how gene expression values look like within a
single sample, we provide gene expression scatter plots for the first
sample in each of the data sets: kidney sample in the Marioni and 2803
sample in the LAML data set (Figure~\ref{fig:absolute_60}). The
microarray gene expression estimates are plotted against
sequencing-based estimates for each of the three RNA-seq data processing
methods: PREBS, MMSEQ and Read counting. In general, the shapes of
scatter plots for all of the methods look similar, however, PREBS
reaches the highest Pearson correlation both on kidney sample in the
Marioni data set ($r = 0.78$) and 2803 sample in the LAML data set ($r
= 0.83$).

We tested the significance of observed correlation differences for a
single sample using r.test() function from psych
package. The significance of difference between PREBS vs microarray
correlation and read counting or MMSEQ vs microarray correlation was
tested taking into account the number of genes for which the
correlation is calculated. All of the observed correlation differences
were significant with \textit{p}-values lower than $10^{-6}$.

\subsection{Retrieval of similar RNA-seq--microarray experiments}

\begin{figure*}[ht] 
 \includegraphics{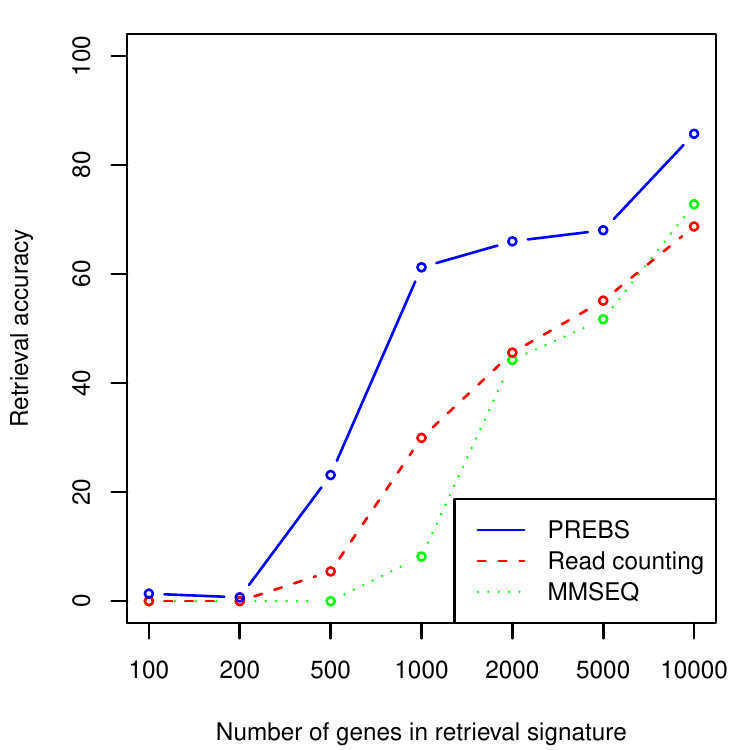}
  \begin{center}
    \caption{{\bf Retrieval accuracy of coupled RNA-seq--microarray experiments.} The plot shows average precision of retrieving the corresponding
      microarray experiment from a large collection based on
      correlation with expression estimates from RNA-seq as a function
      of the number of genes used as the signature. Accuracy is
      measured as a fraction of the samples which have the largest
      correlation with its true pair.}
    \label{fig:retrieval}
  \end{center}
\end{figure*}

One of our main motivations for developing the PREBS method is
information retrieval, where the
aim is to retrieve similar experiments based on the content, i.e. the
signature of expressed genes.
The higher similarity of RNA-seq and microarray data provided by
PREBS processing should allow combining these two types of data more
effectively.
This kind of joint modelling would
significantly increase the utility of methods for content-based
organisation of large gene expression databases such
as that of~\cite{Schmid2012}.

We designed an experiment to see whether the increased absolute gene
expression correlation of PREBS and microarrays can be useful in
a similar RNA-seq--microarray retrieval task. In this experiment we
had several RNA-seq experiments with a matching
microarray experiment that had to be retrieved from a database.
We used the 183 microarray samples in the LAML
data set, 147 of which had a matched RNA-seq pair.
For each RNA-seq experiment
we calculated gene expression estimate correlation with all
microarray experiments. Accuracy was measured by how often the correct
pair had the highest correlation.  Accuracy of retrieval was
calculated for all three RNA-seq data processing methods: PREBS, MMSEQ and
read counting.

To evaluate the performance of the methods using different sized
signatures, we evaluated the performance of the methods
with different numbers of top expressed genes. As we can see in
the results in Figure~\ref{fig:retrieval}, PREBS has clearly a better agreement with microarrays than
the other RNA-seq data processing methods, especially when relatively small subsets
of most highly expressed genes are used as signatures.  Looking this
another way, PREBS can provide similar accuracy with a signature
that is significantly smaller than what is needed by the other
methods, which can provide
significant computational savings in modelling large databases.

\subsection{Differential expression comparison}

\begin{figure*}[ht] 
 \centering 
 \includegraphics[width=\textwidth]{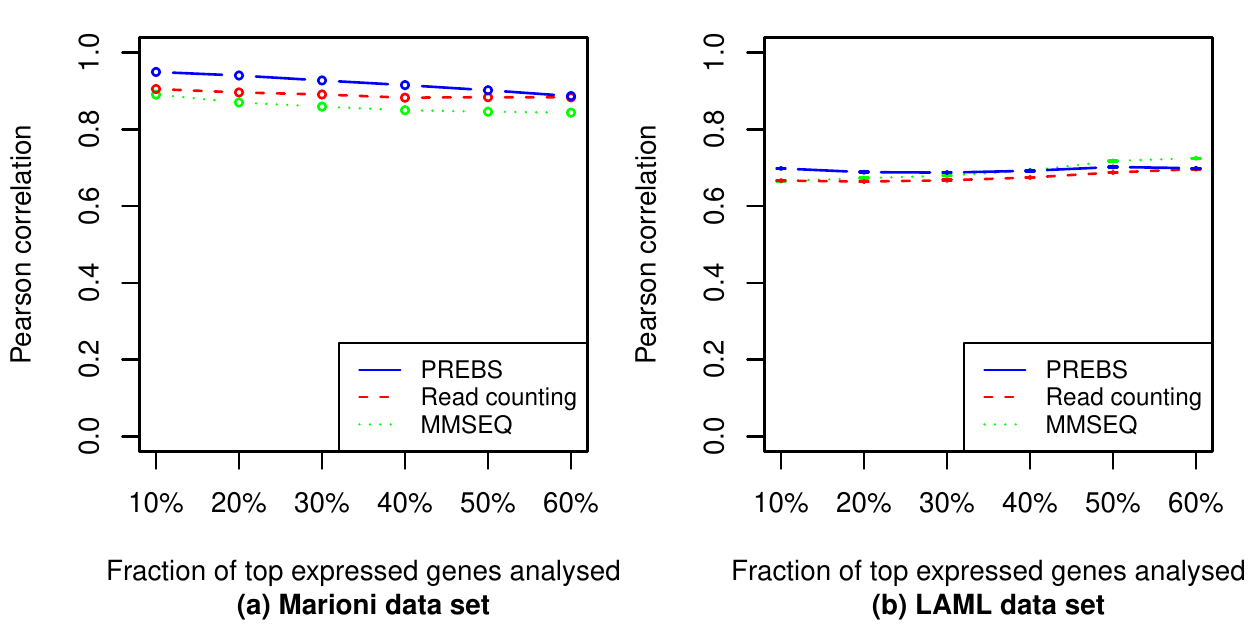}
  
  \caption{{\bf Averaged differential gene expression correlations.} The plots show average $\log_2$ fold change correlations between different RNA-seq
    data processing methods and the microarray. Different points correspond to different numbers of top expressed genes. The correlations are averaged over all samples
    in the corresponding data sets: (a) the Marioni \emph{et al.} data set,
    (b) the LAML data set. The error bars in LAML data set plot
    correspond to standard errors of the mean, although the errors are
    so small that top and bottom bars are merged. Error bars for
    Marioni data set plot could not be displayed because there is only
    one pair of samples for which $\log_2$ fold change values were
    calculated.}
  \label{fig:corplots2}
\end{figure*}

\begin{figure*}[ht] 
 \centering 
 \includegraphics[width=\textwidth]{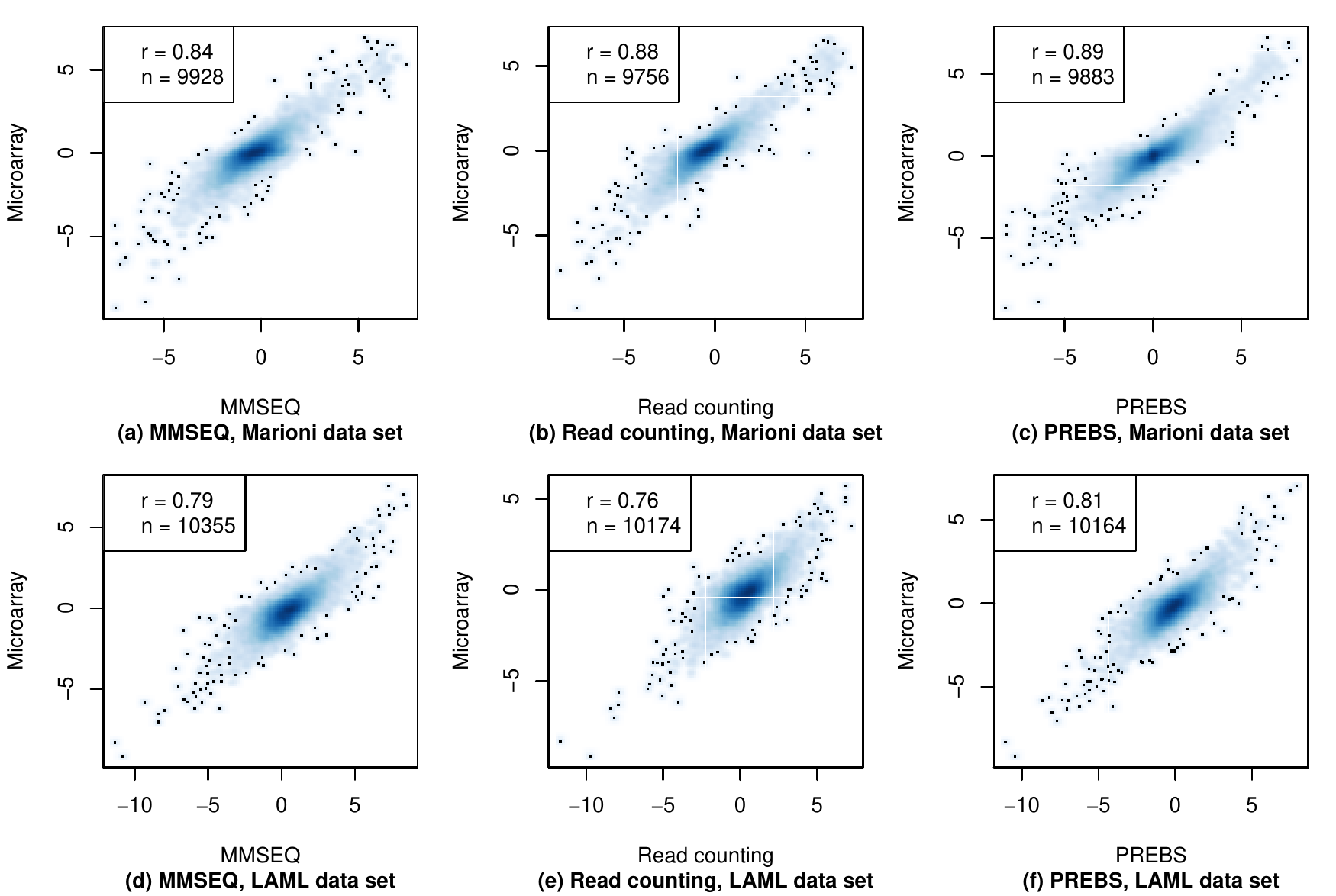}
  
  \caption{{\bf Differential expression scatter plots.} $\log_2$ fold change values for differential expression
    estimated using different RNA-seq analysis methods plotted against
    corresponding microarray $\log_2$ fold change values. The figures
    show 60\% of most highly expressed genes. Only plots for a single sample pair in each data set are shown. The top row shows the fold
    changes between the kidney and liver samples from the Marioni data
    set, while the bottom row shows changes
    between samples 2803 and 2805 from the LAML data set. The legend
    contains Pearson correlation (\textit{r}) and the number of genes (\textit{n}).}
  \label{fig:differential_60}
\end{figure*}

\begin{figure*}[ht] 
 \centering 
 \includegraphics[width=\textwidth]{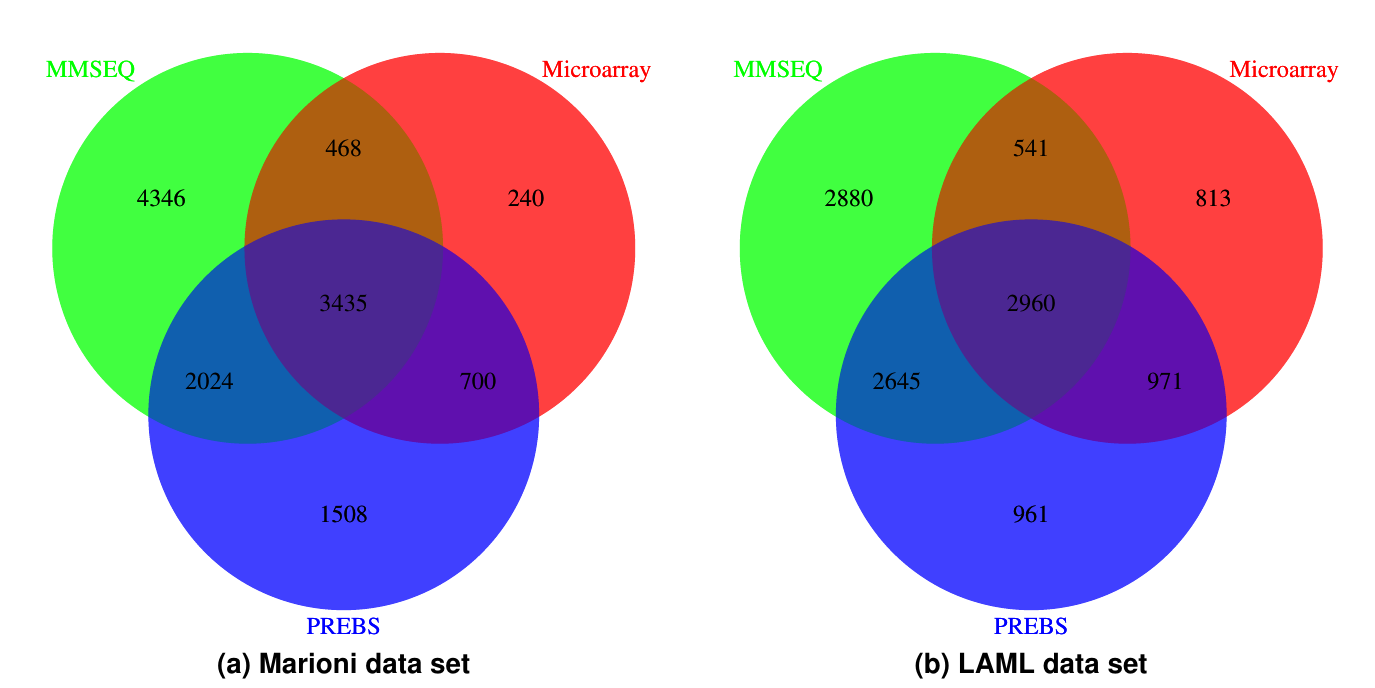}
  
  \caption{{\bf Venn diagrams of differentially expressed genes.} The Venn diagrams illustrate the similarities of lists of
    genes that are called differentially expressed by different methods. We
    call genes with the absolute value of $\log_2$ fold change higher
    than 1.5 as significantly differentially expressed. The pairs of samples that are analyzed are the same as in Figure~\ref{fig:differential_60} (kidney and liver for Marioni data set, 2803 and 2805 for LAML data set).}
  \label{fig:venn}
\end{figure*}

Similarly to the absolute expression comparison, we compared the three
RNA-seq data processing methods based on agreement with microarrays in
differential expression
measurements. For each of the methods and for microarrays we
calculated $\log_2$ fold change values of gene expression between two
states. Our comparison is limited to $\log_2$ fold changes instead
of proper statistical differential expression testing because this
would require biological replicates in RNA-seq data which are not
available in the data sets used, and would be less meaningful anyway
because of the different nature of the tests used on different
platforms.

We evaluated the agreement between the three RNA-seq data processing
methods and microarray by calculating Pearson correlations between
sequencing-based $\log_2$ fold change values and microarray $\log_2$
fold change values. Again we did that for different fractions of top
expressed genes: 10-60\%. Since $\log_2$ fold change calculation
requires two samples, we calculated them for
all possible sample pairs (1 pair for the Marioni data set and
$\binom{147}{2}=10731$ pairs for the LAML data set). So in
Figure~\ref{fig:corplots2} we provide $\log_2$ fold change
correlations averaged over all possible sample pairs in each data set
for different fractions of top expressed genes.

In contrast to the absolute expression case, we see that the differences
in differential expression correlations between different methods are
very small. PREBS method performs slightly better on the higher end of
expression (10-20\%), but slightly worse on the lower end of
expression (50-60\%). We can also see that the differential expression
agreement is better in the Marioni data set where the expression
difference between the samples is large than in the LAML data set where
the samples have quite similar expression levels.

We provide an example of gene expression scatter plots for
differential expression for the first pair of samples of each data set
in Figure~\ref{fig:differential_60}. Again we can see that the shapes of scatter
plots look rather similar between different methods. The correlation
levels differ slightly, but not as much as in absolute expression
case.

Figure~\ref{fig:venn} shows a comparison of the numbers of genes that
have absolute value of $\log_2$ fold change greater than 1.5 (the
criterion for differential expression used e.g.\ in~\cite{beane2011})
for example sample pairs in both data sets. According to these
results, PREBS has a better correlation with microarray results by
having less genes detected alone and having many more
genes in common with microarrays in the LAML data set. MMSEQ finds
more differentially expressed genes than either PREBS or microarray in
both data sets. The added sensitivity arises most likely because it
uses read data from the whole gene regions, while PREBS restricts
itself only to the gene regions where microarray probes are located.
Overall, this again confirms that PREBS results agree with microarray
better than MMSEQ results.

\subsection{Cross-platform differential expression}

\begin{figure*}[ht] 
 \centering 
 \includegraphics[width=\textwidth]{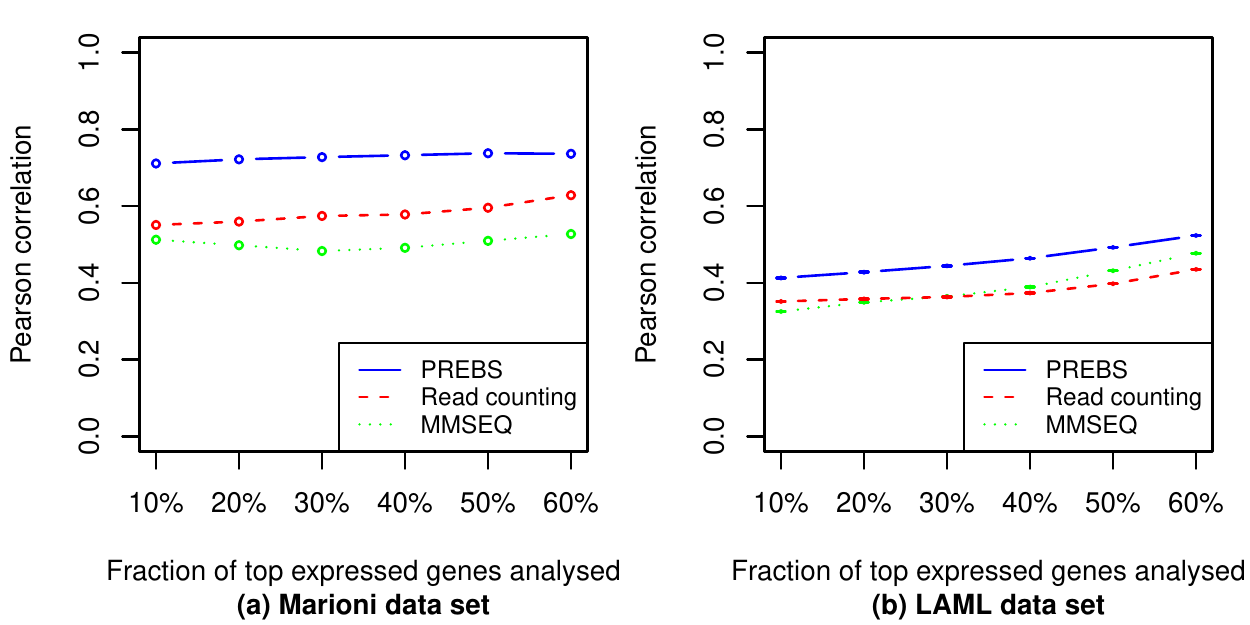}
  
  \caption{{\bf Averaged cross-platform differential gene expression correlations.} The plots show average cross-platform differential gene expression correlations between different RNA-seq
    data processing methods and the microarray. Different points correspond to different numbers of top expressed genes. The correlations are averaged over all possible pairs of samples
    in the corresponding data sets: (a) the Marioni \emph{et al.} data set,
    (b) the LAML data set.}
  \label{fig:corplots3}
\end{figure*}

Better comparability between microarray and RNA-seq data also allows
completely new operations, such as cross-platform differential
expression analysis between samples measured with different
technologies.  This is a very difficult task because RNA-seq and
microarray measures suffer from different biases, and the results
of any such analysis should always be interpreted with care.

To compute the cross-platform differential expression fold change we
perform an extreme quantile normalisation by replacing RNA-seq gene
expression measures with microarray gene expression measures having
corresponding ranks in the coupled experiment. This way, we have not
changed the relative order expression levels, but made the dynamic
ranges of the two platforms identical.

The correlation plots of $\log_2$ fold changes for cross-platform
differential gene expression are shown in Figure~\ref{fig:corplots3}.
We can see that PREBS has significantly better agreement with microarrays than the two other methods
both on Marioni and LAML data sets and can reach a reasonable
level of correlation especially with the Marioni data.  The relative
performances of the different methods mirror those in
Figure~\ref{fig:corplots1} because the performance depends mainly on
similarity of absolute expression measures.

\subsection{Probe set expression calculation}

\begin{figure*}[ht] 
 \centering 
 \includegraphics[width=\textwidth]{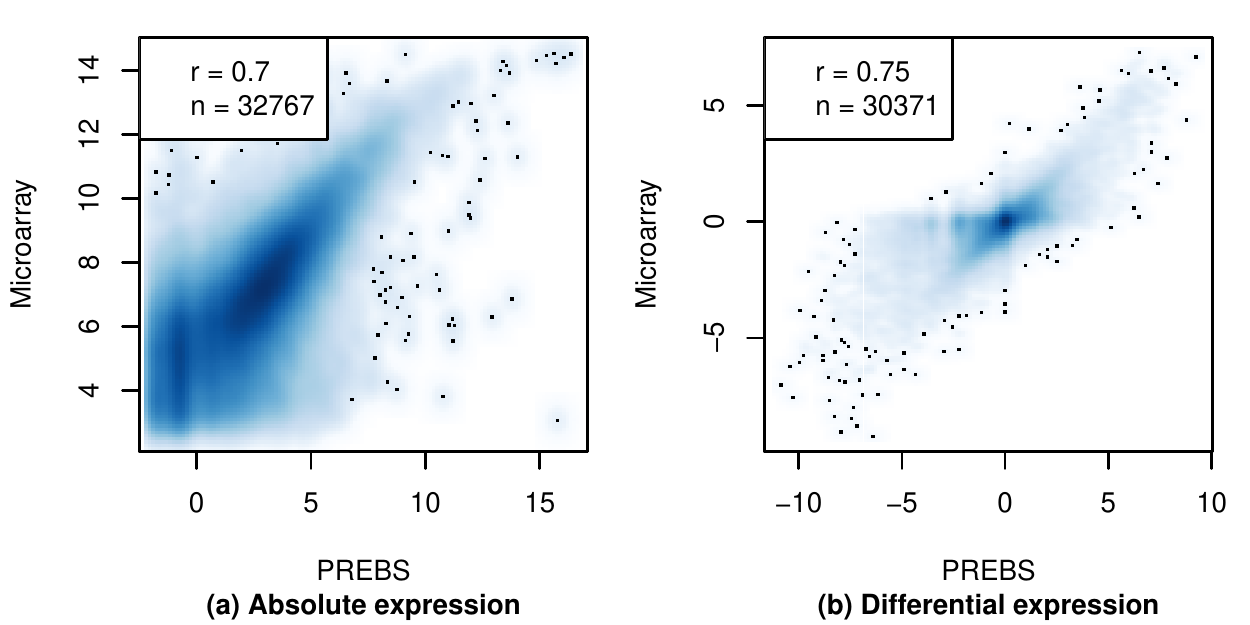}
  
  \caption{{\bf Original microarray probe set gene expression scatter plots.} The plots show (a) estimated absolute expression values and (b) estimated $\log_2$ fold
    changes values for original microarray probe sets. The plots
    show 60\% most highly expressed genes in the Marioni data set.}
  \label{fig:manufacturer}
\end{figure*}

So far we discussed only the results where both PREBS and microarray
were processed using Custom CDF files and gene expression values for
Ensembl gene identifiers were acquired. However, the default way to
process microarray data is using microarray probe set definitions.
PREBS has an option to be run this way too, and in this way it can
produce sequencing-based probe
set expression values that can be directly compared with microarray
probe set expression estimates.

Figure~\ref{fig:manufacturer} shows the scatter plots for absolute and
differential probe set expression estimates using PREBS method on
the Marioni data set. Calculating expression values for probe sets is a
unique feature of PREBS and there is no easy way to do that using
MMSEQ or read counting. Therefore, we did not compare PREBS with these
two methods in this case.

Comparing PREBS vs microarray expression correlations of the two
settings we see that the correlations for manufacturer's probe sets
(Figure~\ref{fig:manufacturer}) are slightly lower than the
correlations for Ensembl genes
(Figure~\ref{fig:absolute_60} and Supplementary Figure 1). However,
this is most likely due to the fact that there are many more probe
sets than genes and the estimation of the corresponding individual
expression levels is less reliable. Overall, PREBS provides a very
reasonable level of correlation with original probe set expression
levels.

\section{DISCUSSION}

Our results clearly demonstrate that the PREBS method is able to
produce from RNA-seq data gene expression estimates that are
significantly more similar to microarray estimates than standard
processing pipelines.  What is more, PREBS
allows obtaining estimates for original microarray probe sets, which
is not possible with existing methods.  This will greatly aid in
building integrated models of large gene expression databases that
contain both microarray and RNA-sequencing data.
These larger databases will help in developing more accurate machine
learning methods for various predictive tasks (e.g.~\cite{Shi2011}).

One potential criticism against the PREBS approach is that it throws
away data in the analysis.  There does not however seem to be an easy
way around this: microarrays only measure the expression of the probe
sequences, and including RNA-seq data over other regions risks
introducing confounding information due to unforeseen splicing and
annotation effects.  It might be possible to develop a more complex
model taking all this into account, but that would be far more
computationally demanding and hence less well-suited for analysis of
large data collections.

PREBS greatly improves the comparability of absolute expression
measures, but it does not provide a significant improvement for
differential expression analysis.  This may in part be explained by
microarray probes that target the gene sequence suboptimally, possibly
focusing only on a small fraction of its alternatively spliced
isoforms.  This introduces a gene-specific bias to the expression
estimates.  When computing the difference between multiple samples,
these biases tend to cancel.  The good performance of PREBS suggests
that focusing on probe regions is likely a significant gene-specific bias in
microarrays.  Learning a model of these and other biases, such as
those caused by different melting points and affinities of the probes,
is an important avenue of future work, but a detailed model will
require a significant amount of diverse paired RNA-seq--microarray
data.

Different experimental techniques for measuring gene expression 
produce different results partly because they measure different 
things, such as different parts of the gene sequence. In this work we 
have presented the PREBS method which aims to eliminate this difference 
from RNA-seq and microarray gene expression analyses by focusing the 
RNA-seq summarisation to microarray probe regions. Combining this with
a standard microarray data processing algorithm leads to estimates
of absolute expression that are significantly more similar to ones measured
from the same samples using microarrays than standard RNA-seq data processing
techniques. The difference between the methods is much smaller in
differential expression, presumably because gene-specific biases
cancel out in the differential analysis.

Diminishing the differences between different gene expression
measurement platforms paves the way for integrative modelling of
large genomic data sets and big genome data applications.
We have demonstrated that the PREBS approach can lead to increased
accuracy in a simplified content-based genomic information retrieval
task. Extending this success to a realistic integrative modelling
system is a very attractive avenue of future research.

\section{ACKNOWLEDGEMENTS}

The authors thankfully acknowledge the TCGA research network for
providing some of the data for this work.
This work was supported by the Academy of Finland [grant number 259440
to A.H.].

\small

\bibliographystyle{myunsrt}
\bibliography{prebs-sources}

\end{document}